\documentclass[%
reprint,
superscriptaddress,
amsmath,amssymb,amsthm,mathrsfs,
aps,
prl,
]{revtex4-1}

\usepackage{graphicx}
\usepackage{dcolumn}
\usepackage{bm}
\usepackage[usenames]{color}
\usepackage[ 
margin=0.75in,
]{geometry}

\def\v2o3{V$_2$O$_3$}
\def\vo{V$_2$O$_3$}

\def\musr{$\mu$SR}
\def\TN{$T_{\mathrm{N}}$}

\newcommand{\ra}[1]{\renewcommand{\arraystretch}{#1}}

\begin{document}

\title{
Intertwined magnetic, structural, and electronic transitions in V$_2$O$_3$
}

\author{Benjamin A. Frandsen}
\affiliation{%
	Department of Physics, Columbia University, New York, NY 10027, USA.
}%
\affiliation{%
	Department of Physics and Astronomy, Brigham Young University, Provo, UT 84602, USA.
}%
\email{benfrandsen@byu.edu}

\author{Yoav Kalcheim}
\affiliation{%
	Department of Physics, University of California San Diego, 9500 Gilman Drive, La Jolla, California 92093, USA.
}%

\author{Ilya Valmianski}
\affiliation{%
	Department of Physics, University of California San Diego, 9500 Gilman Drive, La Jolla, California 92093, USA.
}%

\author{Alexander S. McLeod}
\affiliation{%
	Department of Physics, University of California San Diego, 9500 Gilman Drive, La Jolla, California 92093, USA.
}%

\author{Z. Guguchia}
\affiliation{%
	Department of Physics, Columbia University, New York, NY 10027, USA.
}%
\affiliation{%
	Laboratory for Muon Spin Spectroscopy, Paul Scherrer Institut, CH-5232 Villigen PSI, Switzerland.
}%

\author{Sky C. Cheung}
\affiliation{%
	Department of Physics, Columbia University, New York, NY 10027, USA.
}%

	\author{Alannah M. Hallas}
	\affiliation{%
		Department of Physics and Astronomy, McMaster University, Hamilton, Ontario L8S 4M1, Canada.
	}%

	\author{Murray N. Wilson}
	\affiliation{%
		Department of Physics and Astronomy, McMaster University, Hamilton, Ontario L8S 4M1, Canada.
	}%
	
	\author{Yipeng Cai}
	\affiliation{%
		Department of Physics and Astronomy, McMaster University, Hamilton, Ontario L8S 4M1, Canada.
	}%
	
	\author{Graeme M. Luke}
	\affiliation{%
		Department of Physics and Astronomy, McMaster University, Hamilton, Ontario L8S 4M1, Canada.
	}%
	\affiliation{%
		Canadian Institute for Advanced Research, Toronto, Ontario L8S 4M1, Canada.
	}%

\author{Z. Salman}
\affiliation{%
	Laboratory for Muon Spin Spectroscopy, Paul Scherrer Institut, CH-5232 Villigen PSI, Switzerland.
}%

\author{A. Suter}
\affiliation{%
	Laboratory for Muon Spin Spectroscopy, Paul Scherrer Institut, CH-5232 Villigen PSI, Switzerland.
}%

\author{T. Prokscha}
\affiliation{%
	Laboratory for Muon Spin Spectroscopy, Paul Scherrer Institut, CH-5232 Villigen PSI, Switzerland.
}%
		
	\author{Taito Murakami}
	\affiliation{ %
		Department of Energy and Hydrocarbon Chemistry, Graduate School of Engineering, Kyoto University, Nishikyo, Kyoto 615-8510, Japan.
	} %
	
	\author{Hiroshi Kageyama}
	\affiliation{ %
		Department of Energy and Hydrocarbon Chemistry, Graduate School of Engineering, Kyoto University, Nishikyo, Kyoto 615-8510, Japan.
	} %

\author{D. N. Basov}
\affiliation{%
	Department of Physics, University of California San Diego, 9500 Gilman Drive, La Jolla, California 92093, USA.
}%
\affiliation{%
	Department of Physics, Columbia University, New York, NY 10027, USA.
}%
	
	\author{Ivan K. Schuller}
	\affiliation{%
		Department of Physics, University of California San Diego, 9500 Gilman Drive, La Jolla, California 92093, USA.
	}%
		
	\author{Yasutomo J. Uemura}
	\email{yu2@columbia.edu}
	\affiliation{%
		Department of Physics, Columbia University, New York, NY 10027, USA.
	}%

\begin{abstract}
We present a coordinated study of the paramagnetic-to-antiferromagnetic, rhombohedral-to-monoclinic, and metal-to-insulator transitions in thin-film specimens of the classic Mott insulator \vo\ using low-energy muon spin relaxation, x-ray diffraction, and nanoscale-resolved near-field infrared spectroscopic techniques. The measurements provide a detailed characterization of the thermal evolution of the magnetic, structural, and electronic phase transitions occurring in a wide temperature range, including quantitative measurements of the high- and low-temperature phase fractions for each transition. The results reveal a stable coexistence of the high- and low-temperature phases over a broad temperature range throughout the transition. Careful comparison of temperature dependence of the different measurements, calibrated by the resistance of the sample, demonstrates that the electronic, magnetic, and structural degrees of freedom remain tightly coupled to each other during the transition process. We also find evidence for antiferromagnetic fluctuations in the vicinity of the phase transition, highlighting the important role of the magnetic degree of freedom in the metal-insulator transition.
\end{abstract}

\maketitle

\section{Introduction}

Metal-insulator transitions (MITs) in strongly correlated electron materials are among the most intensely studied topics in condensed matter physics. The interest stems from the intellectual richness of strongly interacting electron systems and the close relationship between MITs and exotic phenomena such as high-temperature superconductivity~\cite{mott;ppsl37,mott;ppsl49,imada;rmp98,basov;rmp11,keime;n15}. Recent work also shows that MITs hold technological promise for hardware-based artificial neural networks~\cite{yang;armr11,picke;nm13,stoli;afm17,yi;nc18,delva;arxiv19}, intensifying the effort to understand the mechanism of the MIT in various strongly correlated materials.

\vo\ is one of the most important materials for studying correlation-induced MITs and the resulting Mott insulating state (used here broadly to refer to an electrically insulating state caused by electron-electron interactions)~\cite{mcwha;prl69,mcwha;prl71,mcwha;prb73}. At ambient pressure, bulk \vo\ undergoes an abrupt MIT when cooled below $\sim$160~K. A structural phase transition from rhombohedral symmetry in the metallic state to monoclinic symmetry in the insulating state~\cite{derni;jpcs70} accompanies the MIT, together with antiferromagnetic (AF) ordering of the vanadium moments~\cite{moon;prl70,uemur;hfi84,denis;jap85,bao;prl93,frand;nc16}. Extensive theoretical studies have contributed significantly to our understanding of this system~\cite{mcwha;prl69,caste;prb78a,caste;prb78b,caste;prb78c,paola;prl99,park;prb00,rozen;prl95,held;prl01,mo;prl03,kelle;prb04,kotli;pht04}, yet the MIT in \vo\ still eludes a comprehensive theoretical description.

A frequent consequence of the coupling among the charge, lattice, and spin degrees of freedom is the spontaneous appearance of inhomogeneous electronic phases~\cite{dagot;s05}. Recently, this type of phase inhomogeneity was observed directly in \vo\ by nanoscale-resolved infrared spectroscopy (nano-IR)~\cite{knoll;n99,qazil;s07,mcleo;np16}. These measurements revealed domains of phase-separated metallic and insulating regions throughout the MIT. This nanoscale electronic inhomogeneity is in line with other recent studies on \vo\ finding phase separation in the charge sector with Cr doping~\cite{lupi;nc10} and the spin sector with pressure~\cite{frand;nc16}. X-ray diffraction studies of \vo\ films likewise reveal phase-separated monoclinic and rhombohedral regions in the vicinity of the transition~\cite{mcleo;np16,kalch;prl19}. 

Related to the observation of intrinsic inhomogeneity across the transition in \vo\ is the crucial issue of whether the electronic, structural, and AF transitions occur simultaneously or can be decoupled from each other. Such decoupling would have significant implications for our understanding of what drives the Mott transition and for potential technological applications. A recent study of \vo\ films combining X-ray diffraction, IR spectroscopy, and transport showed that the electronic and structural transitions are simultaneous~\cite{kalch;prl19}. However, no equivalent work investigating the AF transition in \vo\ films has been reported.

Here, we present low-energy muon spin relaxation/rotation (\musr) experiments on films of \vo\ to determine quantitatively the magnetically ordered phase fraction across the transition, thereby filling the gap left by the earlier works. We also performed temperature-dependent x-ray diffraction (XRD) and nano-IR measurements on the same specimens, offering a uniquely complete picture of the MIT in \vo. Using the sample resistance to calibrate the temperature between the \musr\ and XRD instruments, we quantitatively compare the magnetic and structural phase fractions across the transition. The midpoints of the magnetic and structural transitions are indistinguishable within the experimental sensitivity of $\sim$1~K, but magnetic correlations develop in a partial volume fraction beginning $\sim$10~K above the onset of the structural transition. This suggests that antiferromagnetism plays a dominant role in the MIT.

\section{Experimental Details}

Two sets of \vo\ films of thickness 300~nm and 70~nm, respectively, were grown on a sapphire substrate cut so that the (012) plane is in the out-of-plane direction of the film. We followed the procedure in Ref.~\onlinecite{trast;jms18}, except that the films in the present work were pristine, not annealed. These samples were characterized with low-energy \musr\ at the Paul Scherrer Institut. To prepare each sample for the \musr\ measurements, six identical films of dimensions 12~mm $\times$ 7~mm were tiled together and adhered to a nickel plate, which was inserted in the \musr\ spectrometer on the LEM (Low Energy Muon) beamline at the Paul Scherrer Institute~\cite{proks;nima08}. A pseudo-continuous beam of 100\% spin-polarized muons with initial energies of 4~MeV was slowed by a moderator to energies of about 15~eV, then electrostatically accelerated to a desired energy on the order of 10~keV, allowing implantation of the muons selectively at a variable mean depth ranging from 10~nm to several tens of nm.

In a \musr\ experiment, implanted muons will stop at a site in the crystal corresponding to an electrostatic potential minimum. The spin of each muon precesses around the magnetic field at this site until the muon decays with a mean lifetime of 2.2~$\mu$s into two neutrinos and a positron, the latter being emitted preferentially along the direction of the muon spin at the moment of decay. The normalized difference in positron counts between pairs of opposing positron detectors near the sample is known as the \musr\ asymmetry and is proportional to the projection of the muon ensemble spin polarization along the axis defined by the detector pair. Tracking the asymmetry as a function of time after muon implantation therefore provides detailed information about the local magnetic field distribution in the sample.

Integrating the muon beam intensity profile over the sample area indicated that for both samples, 85\% of the muons landed in the sample area and the remaining 15\% in the nickel plate, which contributes a featureless exponential decay that disappears well before 0.1~$\mu$s in the asymmetry spectra~\cite{saada;pp12}. Of muons landing in the sample area, 10~$\pm$~3.5\% landed in exposed areas of the sapphire substrate, contributing a small and extremely rapid decay that manifests only as absent asymmetry within the time-binning used for the data analysis~\cite{proks;prl07,krieg;prb17}. Therefore, the dominant features observed are intrinsic to the \vo\ films. Least-squares fits to the \musr\ data were performed using the program MUSRFIT~\cite{suter;physproc12}.

The same samples used for the \musr\ experiments were also characterized with XRD using a Rigaku Smartlab instrument equipped with a variable temperature stage as described elsewhere~\cite{kalch;prl19}.  The temperatures measured during the \musr\ and XRD experiments on the 300-nm sample were precisely calibrated to each other in the following way. \textit{In situ} measurements of the resistance $R$ were performed during the \musr\ and XRD experiments, and the measured values were compared to precise $R$ versus $T$ measurements conducted on a separate probe station. The same contact points on the film were used in each case. To account for different contact resistances, the $R$ values measured during the \musr\ and XRD experiments were corrected by adding a constant value such that the minimum resistance (i.e., the resistance in the metallic state at a temperature just above the onset of the MIT) agreed with the minimum resistance measured on the probe station. This offset was 9.8~$\Omega$ for the \musr\ measurements and 2.4~$\Omega$ for XRD. For a given (corrected) value of $R$ measured during the \musr\ or XRD experiments, the calibrated temperature was taken to be the corresponding temperature measured on the probe station. This procedure was not performed during any of the experiments on the 70-nm sample, so we regard the temperatures reported for that sample to be unsuitable for quantitative comparison among the different experimental techniques. In addition to the \musr\ and XRD characterization, the 70-nm sample was further studied by nano-IR with a spatial resolution of $\sim$25~nm using methods described previously~\cite{mcleo;np16}. 

\section{Results and Discussion}

\subsection{\musr\ Characterization}
\begin{figure}
	
	\includegraphics[width=80mm]{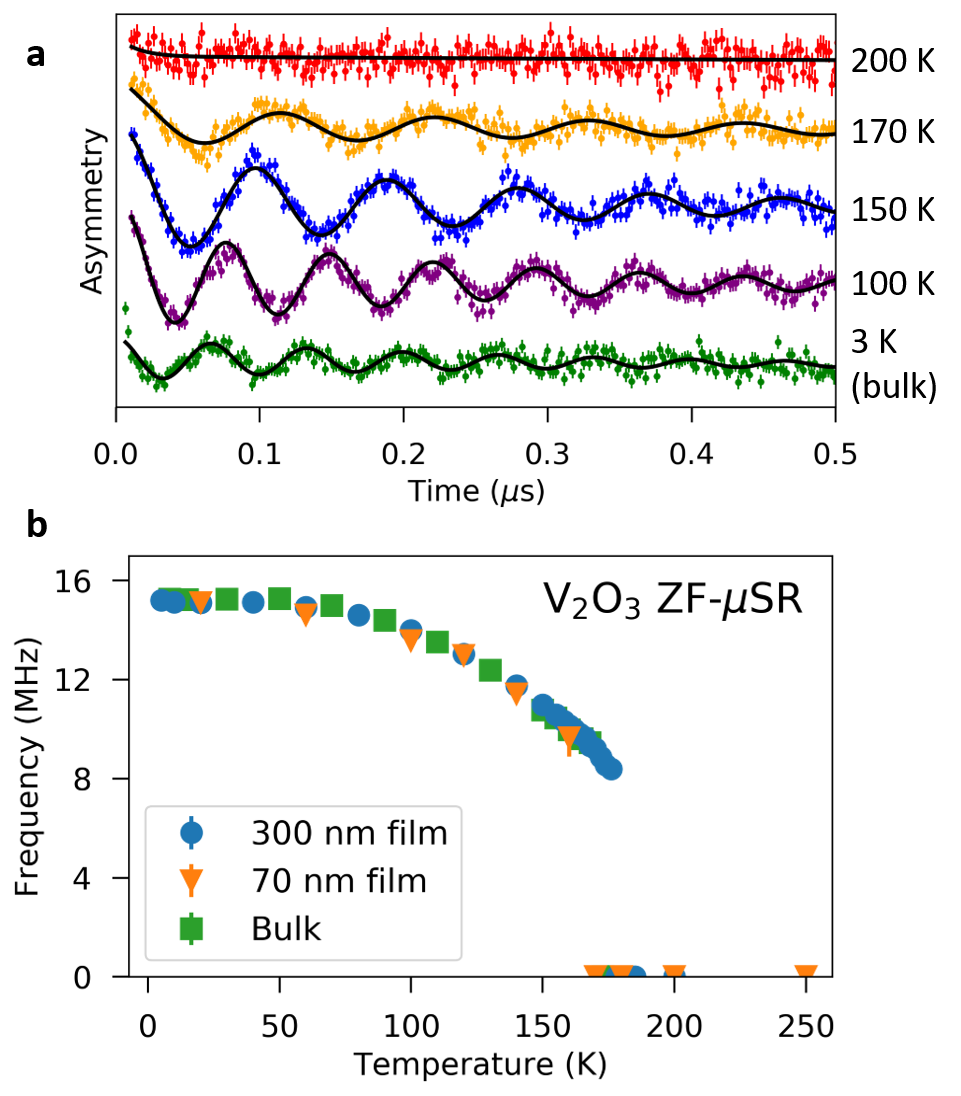}
	\caption{\label{fig:zf} (Color online) (a) ZF \musr\ time spectra from the 300-nm sample measured in a warming sequence spanning the antiferromagnetic ordering temperature, offset vertically for clarity. A spectrum collected from bulk \v2o3\ is also shown. (b) Oscillation frequencies for the 300-nm, 70-nm, and bulk samples of \vo\ extracted from least-squares refinements performed against the ZF spectra. Error bars represent the estimated standard deviation (ESD) of the frequency determined by the refinements. Note that the 70-nm film data were collected on cooling and the others on warming.}
	
\end{figure}
We first present \musr\ measurements of the 300-nm sample in zero external field (ZF) to probe the intrinsic AF order in the sample. In Fig.~\ref{fig:zf}(a), we display representative ZF spectra measured with 15-keV muons, corresponding to a mean stopping depth of $70 \pm 20$~nm. The overlaid black curves show fits using a model consisting of an exponential component for the background from the nickel mounting plate (visible as the slight decay at early time for the 200~K spectrum), an exponentially damped cosine function for the oscillations of the muon spin component transverse to the local magnetic field, and an exponential component for the spin component longitudinal to the local field. Above the AF ordering temperature \TN, the asymmetry contribution from \vo\ relaxes slowly with time (too slowly to be observed in Fig.~\ref{fig:zf}(a)), as expected for a paramagnet. Coherent oscillations develop below \TN, reflecting the establishment of long-range magnetic order. These oscillations are well described by a single oscillating frequency, which we extracted from least-squares fits and display as blue circles in Fig.~\ref{fig:zf}(b). This frequency is proportional to the magnitude of the staggered AF moment. Upon warming, the frequency decreases steadily until oscillations abruptly disappear at 176~K. Equivalent analysis on the 70-nm sample with 8-keV muons (stopping depth of $40 \pm 20$~nm) yields very similar results (orange triangles in Fig.~\ref{fig:zf}(b)). For comparison, we also display the ZF spectra of bulk \vo\ measured at 3~K at the TRIUMF \musr\ facility at the bottom of Fig.~\ref{fig:zf}(a) and the corresponding oscillation frequencies for the bulk sample as green squares in panel (b). The bulk and film samples display nearly identical behavior, indicating they share the same magnetic ground state. Further measurements using varying muon energies indicate there is no significant depth-dependence to the magnetism [see the Supplemental Material~\footnote{See Supplemental Material for more information about the \musr\ experiments and the nanoscale-resolved infrared spectroscopy measurements.}]. The substantial reduction of the oscillation frequency when approaching \TN\ from below agrees with previous work~\cite{uemur;hfi84}.

To quantify the temperature dependence of the magnetically ordered phase fraction in the 300-nm sample, we performed measurements in a weak field ($\sim$50~G) directed transverse to the initial muon spin (wTF configuration) using 15~keV muons. The spins of muons landing in paramagnetic regions of the sample exhibit slow precession around the weak external field, while the spins of muons landing in magnetically ordered regions precess rapidly around the vector sum of the small external field and the much larger local intrinsic field ($\sim$~1.1~kG at 5~K). Within the longer time window used to view and analyze wTF data, the signal from muons landing in ordered regions appears as missing initial asymmetry or a non-oscillating component. Representative wTF spectra are shown in the Supplemental Material. Comparing the temperature-dependent amplitude $a(T)$ of the slow oscillations arising from muons in paramagnetic regions to the measured initial total asymmetry well above the transition $a_{\mathrm{max}}$, the paramagnetic phase fraction can be determined as $[a(T)-a_{\mathrm{min}}]/(a_{\mathrm{max}}-a_{\mathrm{min}})$. Here, $a_{\mathrm{min}}$ is the minimum amplitude in the fully AF state and is very close to zero ($\sim$0.01). This asymmetry is within the typical background contribution at the low-energy \musr\ spectrometer used in this study. However, other contributions from minute non-magnetic regions in the sample cannot be ruled out, such as very thin amorphous phases known to form on the surface. 

\begin{figure}
	
	\includegraphics[width=80mm]{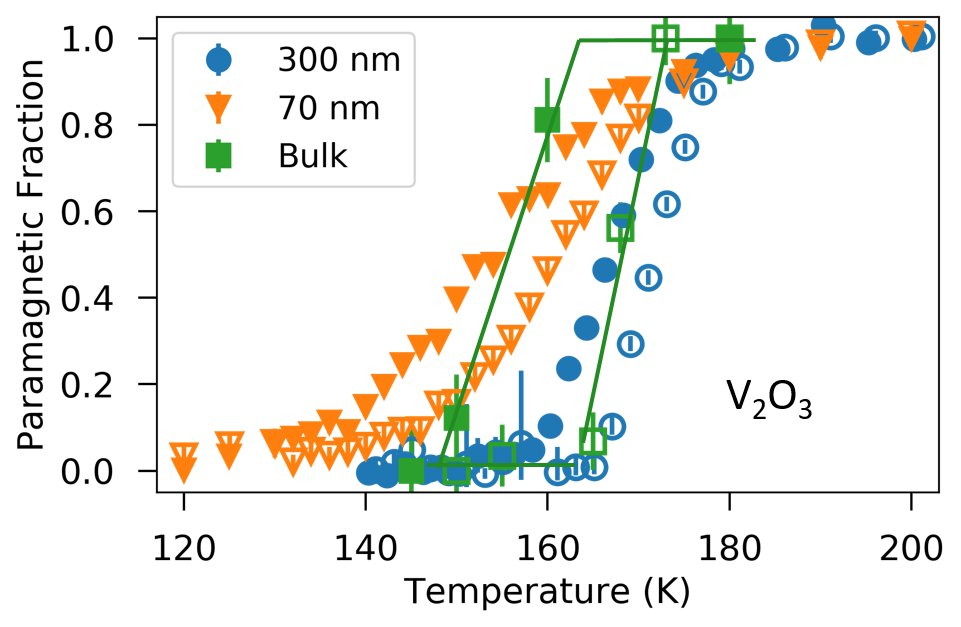}
	\caption{\label{fig:wtf} (Color online) Temperature evolution of the paramagnetic phase fraction in \vo\ for the 300-nm, 70-nm, and bulk samples. Filled symbols represent data taken in a cooling sequence, open symbols in a warming sequence.}
	
\end{figure}

The paramagnetic phase fraction is plotted for the 300-nm sample (blue circles) and 70-nm sample (orange triangles) in Fig.~\ref{fig:wtf}. Open (filled) symbols represent data collected on warming (cooling). Although the onset temperature is similar for the two samples (roughly 175~K), the transition progresses more rapidly with temperature in the 300-nm sample, completing at $\sim$~155~K compared to $\sim$~120~K for the 70-nm sample. The paramagnetic fraction of bulk \vo\ determined from earlier \musr\ measurements~\cite{uemur;hfi84} is shown for comparison (green squares). The bulk transition is sharper than that in either of the film samples, completing within a temperature range of $\sim$~15~K, and exhibits a somewhat wider thermal hysteresis, consistent with other neutron and muon studies~\cite{moon;prl70,blago;prb10}.

These results reveal a relatively slow development of AF order across the transition in the film samples. In the broad temperature interval over which the transition occurs, the magnetic state of the system is inhomogeneous, separated into paramagnetic regions and AF regions. This is similar to the separation into rhombohedral/monoclinic regions and metallic/insulating regions reported in previous studies of \vo\ films~\cite{mcleo;np16,kalch;prl19}.

\subsection{Quantitative Comparison of Magnetic and Structural Phase Fractions}

\begin{figure}
	
	\includegraphics[width=80mm]{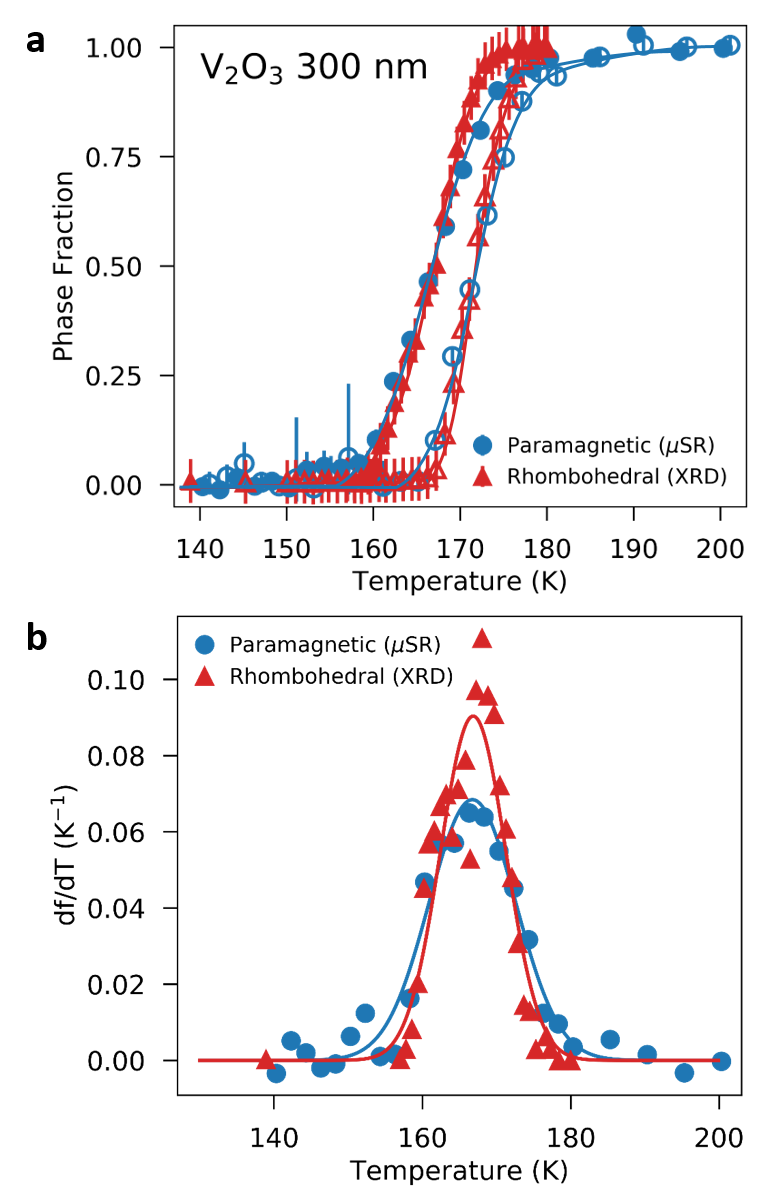}
	\caption{\label{fig:phasefrac300nm} (Color online) (a) Paramagnetic and rhombohedral phase fractions for the 300-nm film. Filled (open) symbols represent data collected on cooling (warming). (b) Temperature derivative of the phase fractions plotted in (a), with Gaussian fits to the data displayed as solid curves.}
	
\end{figure}

We now present a quantitative, temperature-calibrated comparison of the thermal evolution of the paramagnetic and rhombohedral phase fractions of the 300-nm sample to determine whether the two transitions occur simultaneously or separately. XRD measurements were performed between 180~K and 140~K in both cooling and warming sequences. The structural transition causes the (012) rhombohedral Bragg peak to transform into the (011) monoclinic peak, which are separated enough in reciprocal space to be easily distinguished~\cite{kalch;prl19}. In the XRD patterns collected at intermediate temperatures across the transition, both peaks are clearly observable, indicating coexistence of the two phases. To quantify the rhombohedral phase fraction, Lorentzian and Gaussian curves were fit to the (012) rhombohedral and (011) monoclinic Bragg peaks and their relative weights compared~\cite{kalch;prl19}. The uncertainty of this calculation is approximately 10\%~\cite{saerb;jmr14}. In Fig.~\ref{fig:phasefrac300nm}(a), we display the rhombohedral and paramagnetic phase fractions with red triangles and blue circles, respectively.

The paramagnetic and rhombohedral phase fractions show a very similar temperature evolution on both warming and cooling. Within an uncertainty of $\sim$1~K, the temperature $T_{1/2}$ at which the high temperature phase occupies half the sample volume (i.e., the center of the transition) is identical for both the magnetic and structural transitions (167~K on cooling, 171~K on warming). The completion temperatures of the two transitions are likewise very similar. One significant difference between the two transitions is that the temperature at which the paramagnetic fraction begins to drop below unity ($\sim$~185~K) is higher than that of the rhombohedral fraction ($\sim$~175~K). This observation will be discussed subsequently.

Taking a more quantitative approach, we display the numerical derivative $\mathrm{d}f/\mathrm{d}T$ of the high-temperature phase fraction $f$ with respect to temperature $T$ in Fig.~\ref{fig:phasefrac300nm}(b). For simplicity, only the data collected on cooling are considered. We fit a normalized Gaussian model of the form $\mathrm{d}f/\mathrm{d}T = \frac{1}{\sigma \sqrt{2\pi}}\exp\left[-\frac{(T-T_0)^2}{2\sigma^2}\right]$ to the data, with the best fits displayed as solid curves in Fig.~\ref{fig:phasefrac300nm}(b). Table~\ref{table:params} compares the refined parameters. The fitted values of $T_0$ (the temperature at which the phase fraction changes most rapidly) are identical for the two transitions within the fit uncertainty of $~\sim$~0.3~K, and also agree closely with the measured $T_{1/2} = 167$~K. This confirms that the two transitions remain tightly coupled in temperature. For the magnetic transition, the width $\sigma$ is somewhat larger than for the structural transition, which is related to the higher onset temperature in the \musr\ data mentioned previously.

\begin{table}[ht]
	\ra{1.3}
	\caption{Refined Gaussian parameters of the temperature derivative of the low-temperature magnetic and structural phase fractions.} 
	\centering 
	\begin{tabular}{l c c} 
		& $T_0$ (K) & $\sigma$ (K) \\  
		\cline{2-3} 
		\hline
		Magnetic & 166.7 $\pm$ 0.2 & 5.8 $\pm$ 0.2  \\
		Structural & 166.8 $\pm$ 0.3 & 4.4 $\pm$ 0.2  \\
		\hline
	\end{tabular}
	\label{table:params} 
\end{table}


This quantitative comparison of the magnetic and structural transitions in \vo\ films fills an important gap in the existing body of experimental work and provides strong evidence that the magnetic and structural transitions remain tightly coupled. Together with the recent report of robust coupling of the structural and electronic transitions in identically prepared films~\cite{kalch;prl19}, these experiments indicate that the transition in \vo\ occurs simultaneously in the magnetic, structural, and electronic sectors. The magnetic and structural phase fractions of the 70-nm sample were also measured and found to be similar to the 300-nm sample, including an apparently higher-temperature onset of the magnetic transition than the structural transition (see Fig.~\ref{fig:nanoIR}(b)), but the lack of precise temperature calibration between the measurements prevents a more quantitative comparison.

The observation that the paramagnetic fraction begins to decrease at a higher temperature than does the rhombohedral phase fraction merits further discussion. This suggests that quasi-static magnetic correlations develop in a partial volume fraction above the structural phase transition, causing the paramagnetic volume to drop below unity. We expect these to be short-range correlations, since coherent ZF oscillations indicative of genuine long-range magnetic order do not appear until lower temperatures. Given earlier neutron scattering data demonstrating the existence of short-range critical AF fluctuations above \TN~\cite{bao;prl97,bao;prb98}, this gradual reduction of the paramagnetic fraction at high temperature is likely due to partial pinning of these AF fluctuations. Such fluctuations would be expected for a continuous phase transition, despite the fact that the AF transition in \vo\ is first order. This suggests that the critical point of a continuous transition is located nearby in parameter space. Similarly, we note that the ZF precession frequency displayed in Fig.~\ref{fig:zf}(b) shows a pronounced reduction when approaching the transition temperature from below (which would be expected for a continuous magnetic phase transition), but the smooth evolution is ultimately interrupted in first-order-like fashion at a temperature close to the structural phase transition. This is further evidence that the magnetic transition is nearly continuous but is rendered first-order by the structural phase transition. Such a situation is also found in the perovskite rare-earth nickelate Mott systems~\cite{frand;nc16}. We note that these results support recent magnetotransport data and theoretical calculations~\cite{trast;arxiv18}, which suggest that magnetism plays the dominant role in \vo\ through a Mott-Slater mechanism for the MIT. 

\subsection{nanoIR Characterization}
\begin{figure}
	\includegraphics[width=65mm]{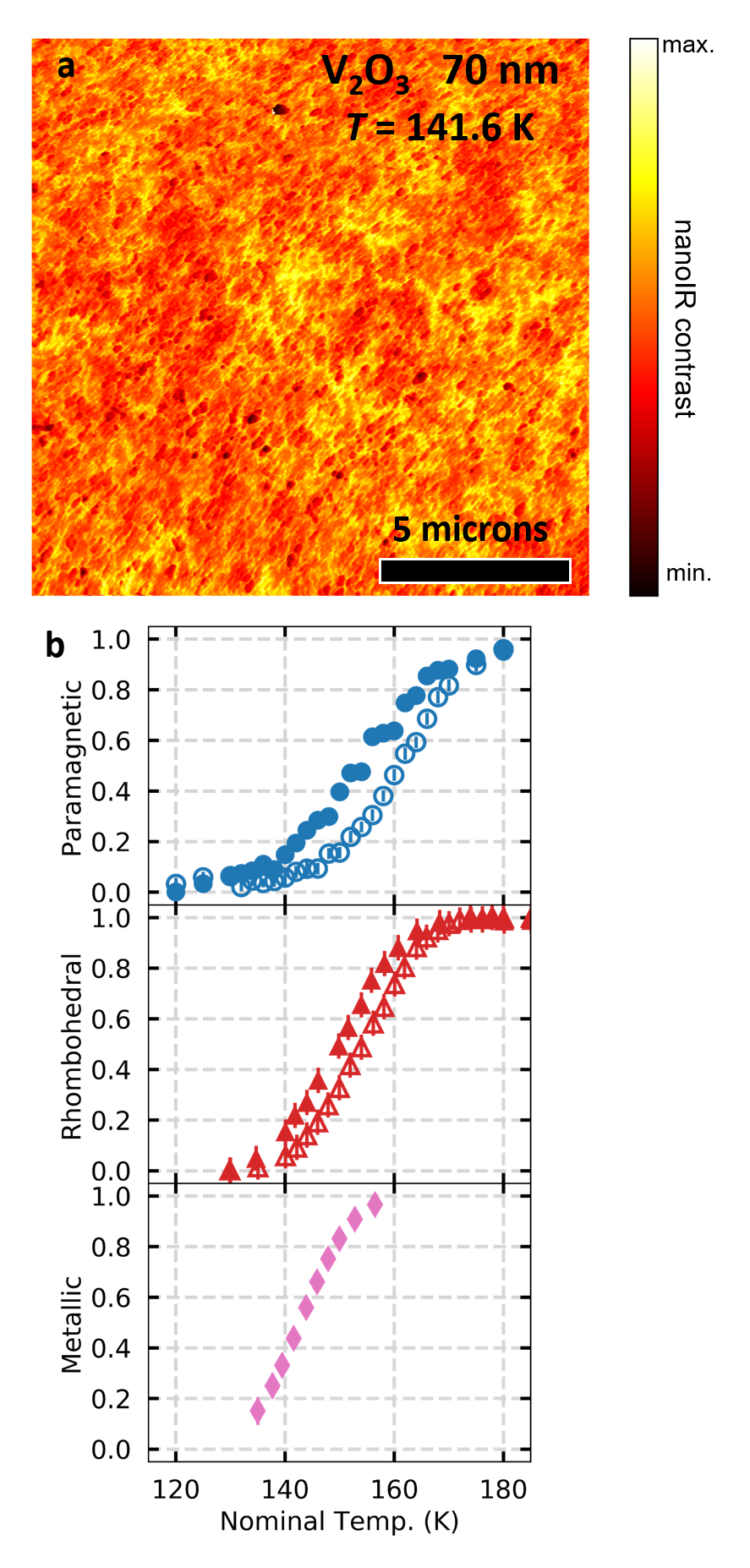}
	\caption{\label{fig:nanoIR} (Color online) (a) Nano-IR measurements of the 70-nm sample of \vo\ at a nominal temperature of 141.6~K showing spatial phase separation between metallic (light-colored) and insulating (dark-colored) domains. (b) Paramagnetic, rhombohedral, and metallic phase fractions for the 70~nm film as a function of nominal temperature (note that the thermometry was not calibrated between experimental techniques). Filled (open) symbols represent data collected on cooling (warming).}	
\end{figure}

Having examined the magnetic and structural transitions, we complete our study by probing the MIT through nano-IR measurements of the 70-nm sample. Spatially resolved maps of the local IR response were generated at several temperatures spanning the transition. In Fig.~\ref{fig:nanoIR}(a), we display a representative image taken at a nominal temperature of 141.6~K. Micron-scale metallic and insulating domains (light and dark regions, respectively) coexist in a spatially segregated manner, confirming that widespread electronic inhomogeneity across the transition accompanies the magnetic and structural inhomogeneity described previously.

These maps of the local IR response can be analyzed to estimate the metallic and insulating fractions for comparison with the magnetic and structural phase fractions. Details of this procedure are provided in the SI. In Fig.~\ref{fig:nanoIR}(b), we display the metallic phase fraction together with the previously described paramagnetic and rhombohedral phase fractions determined for the same 70-nm sample. The temperature evolution of all three transitions is qualitatively similar, exhibiting a gradual development of the low-temperature phase over a $\sim$~30~K interval centered roughly around 150~K. A quantitative comparison is not possible due to the absence of \textit{in-situ} resistance measurements to enable accurate temperature calibration across the three experimental setups. The apparent differences in the midpoint transition temperature of the magnetic, structural, and electronic transitions evident in Fig.~\ref{fig:nanoIR}(b) should therefore be treated cautiously. Nevertheless, it is clear that real-space phase separation between the low- and high-temperature states occurs over a similar temperature interval in all three sectors. The application of these three complementary techniques to the same 70-nm specimen represents a uniquely complete characterization of the magnetic, structural, and electronic phase transitions in \vo.


\section{Summary}
We have reported complementary \musr, XRD, and nano-IR measurements of two samples of \vo\ in thin-film form. The experimental results reveal real-space phase segregation between the low- and high-temperature phases throughout the magnetic, structural, and electronic transitions. Furthermore, we presented a quantitative comparison of the antiferromagnetic and monoclinic phase fractions with precisely calibrated thermometry, enabling an unambiguous comparison of the magnetic and structural phase evolution. We found that the midpoints of the two transitions occur simultaneously within our experimental sensitivity of $\sim$~1~K, consistent with the simultaneous structural and electronic transitions reported previously~\cite{kalch;prl19}. Together, these results demonstrate that the spin, lattice, and charge sectors remain tightly coupled across the Mott transition in \vo. However, we also observe evidence for AF fluctuations at temperatures both above and below the structural phase transition. At low temperature, this manifests as a smooth reduction of the ZF-\musr\ oscillation frequency; at high temperature, as a reduction of the paramagnetic phase fraction with no observable ZF oscillations. This suggests that both the low- and high-temperature phases of \vo\ are close to continuous AF phase transitions which are preempted by a structural transition, rendering the MIT first order. These findings suggest that magnetism plays an important role in the MIT, in agreement with recent experimental and theoretical work~\cite{trast;arxiv18}.

\begin{acknowledgments}

	\textbf{Acknowledgments.} This is a highly collaborative research effort. Experiments were designed jointly and the manuscript was written in multiple iterations by all coauthors. This work is based on experiments performed at the Swiss Muon Source S$\mu$S, Paul Scherrer Institute, Villigen, Switzerland. Synthesis, structure and transport measurements were performed at UCSD (Y.K., I.V. and I.K.S) under grant DE-FG02-87ER-45322. Work at Columbia University was supported by the U.S. National Science Foundation (NSF) via Grant DMREF DMR-1436095, NSF Grant No. DMR-1105961, NSF Grant No. DMR-1610633, and the NSF PIRE program through Grant No. OISE-0968226, with additional support from the Japan Atomic Energy Agency Reimei Project and the Friends of Todai Foundation. BAF acknowledges support from the NSF GRFP under Grant No. DGE-11-44155 and support from the College of Physical and Mathematical Sciences at Brigham Young Univeresity. A.S.M. and D.N.B. are supported by ARO grant W911NF-17-1-0543. Development of nano-optical instrumentation at Columbia is supported by AFOSR FA9550-15-1-0478. Work at Kyoto University was supported by JSPS Core-to-Core Program (a) Advanced Research Networks.

\end{acknowledgments}

\end{document}


\title{
Supplementary Information: Intertwined magnetic, structural, and electronic transitions in V$_2$O$_3$
}

\author{Benjamin A. Frandsen}
\affiliation{%
	Department of Physics, Columbia University, New York, NY 10027, USA.
}%
\affiliation{%
	Department of Physics and Astronomy, Brigham Young University, Provo, UT 84602, USA.
}%
\email{benfrandsen@byu.edu}

\author{Yoav Kalcheim}
\affiliation{%
	Department of Physics, University of California San Diego, 9500 Gilman Drive, La Jolla, California 92093, USA.
}%

\author{Ilya Valmianski}
\affiliation{%
	Department of Physics, University of California San Diego, 9500 Gilman Drive, La Jolla, California 92093, USA.
}%

\author{Alexander S. McLeod}
\affiliation{%
	Department of Physics, University of California San Diego, 9500 Gilman Drive, La Jolla, California 92093, USA.
}%

\author{Z. Guguchia}
\affiliation{%
	Department of Physics, Columbia University, New York, NY 10027, USA.
}%
\affiliation{%
	Laboratory for Muon Spin Spectroscopy, Paul Scherrer Institut, CH-5232 Villigen PSI, Switzerland.
}%

\author{Sky C. Cheung}
\affiliation{%
	Department of Physics, Columbia University, New York, NY 10027, USA.
}%

	\author{Alannah M. Hallas}
	\affiliation{%
		Department of Physics and Astronomy, McMaster University, Hamilton, Ontario L8S 4M1, Canada.
	}%

	\author{Murray N. Wilson}
	\affiliation{%
		Department of Physics and Astronomy, McMaster University, Hamilton, Ontario L8S 4M1, Canada.
	}%
	
	\author{Yipeng Cai}
	\affiliation{%
		Department of Physics and Astronomy, McMaster University, Hamilton, Ontario L8S 4M1, Canada.
	}%
	
	\author{Graeme M. Luke}
	\affiliation{%
		Department of Physics and Astronomy, McMaster University, Hamilton, Ontario L8S 4M1, Canada.
	}%
	\affiliation{%
		Canadian Institute for Advanced Research, Toronto, Ontario L8S 4M1, Canada.
	}%

\author{Z. Salman}
\affiliation{%
	Laboratory for Muon Spin Spectroscopy, Paul Scherrer Institut, CH-5232 Villigen PSI, Switzerland.
}%

\author{A. Suter}
\affiliation{%
	Laboratory for Muon Spin Spectroscopy, Paul Scherrer Institut, CH-5232 Villigen PSI, Switzerland.
}%

\author{T. Prokscha}
\affiliation{%
	Laboratory for Muon Spin Spectroscopy, Paul Scherrer Institut, CH-5232 Villigen PSI, Switzerland.
}%
		
	\author{Taito Murakami}
	\affiliation{ %
		Department of Energy and Hydrocarbon Chemistry, Graduate School of Engineering, Kyoto University, Nishikyo, Kyoto 615-8510, Japan.
	} %
	
	\author{Hiroshi Kageyama}
	\affiliation{ %
		Department of Energy and Hydrocarbon Chemistry, Graduate School of Engineering, Kyoto University, Nishikyo, Kyoto 615-8510, Japan.
	} %

\author{D. N. Basov}
\affiliation{%
	Department of Physics, University of California San Diego, 9500 Gilman Drive, La Jolla, California 92093, USA.
}%
\affiliation{%
	Department of Physics, Columbia University, New York, NY 10027, USA.
}%
	
	\author{Ivan K. Schuller}
	\affiliation{%
		Department of Physics, University of California San Diego, 9500 Gilman Drive, La Jolla, California 92093, USA.
	}%
		
	\author{Yasutomo J. Uemura}
	\email{yu2@columbia.edu}
	\affiliation{%
		Department of Physics, Columbia University, New York, NY 10027, USA.
	}%

\maketitle

\begin{figure}
	
	\includegraphics[width=80mm]{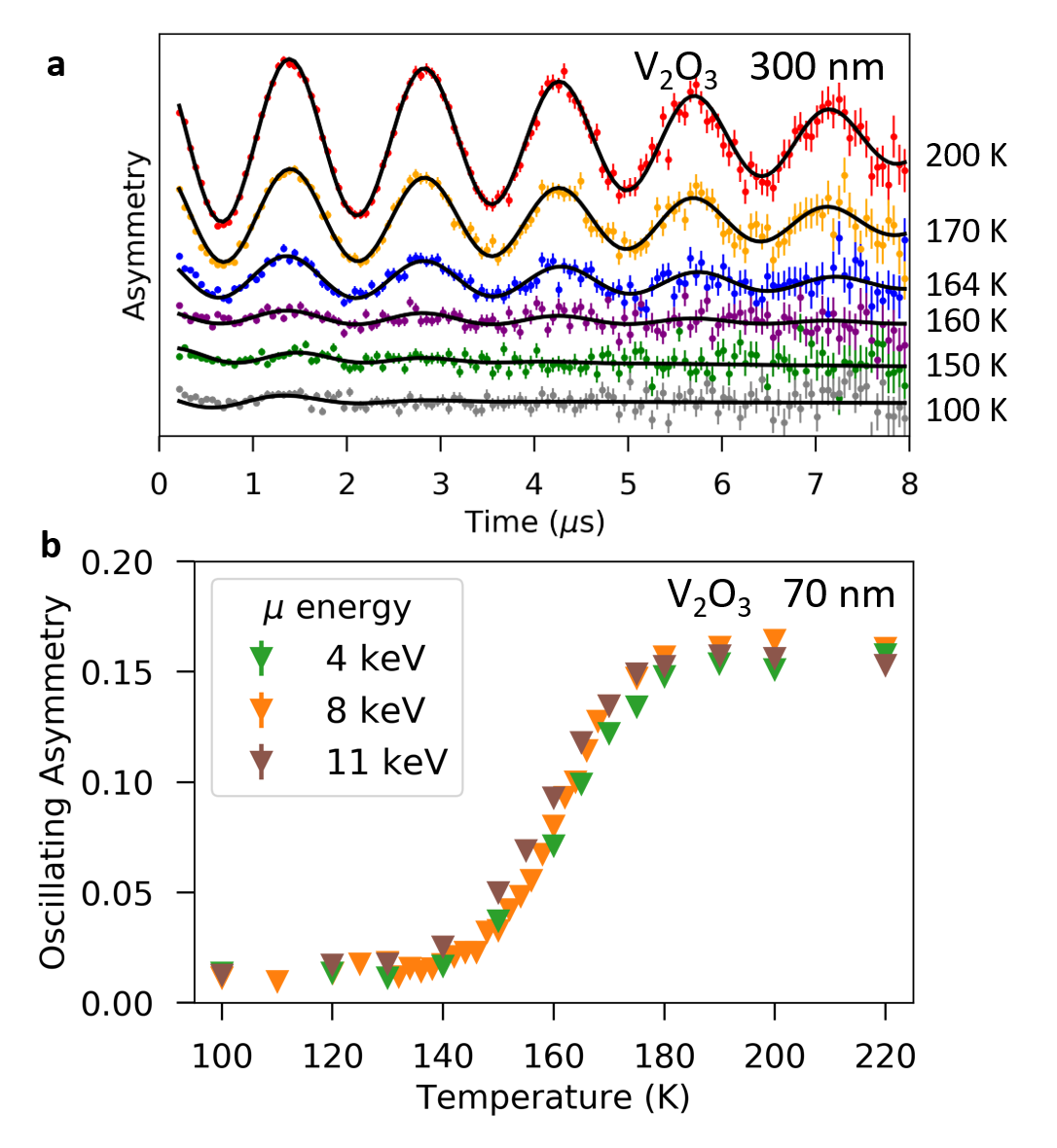}
	\caption{\label{fig:wtf} (a) Representative wTF \musr\ spectra collected from the 300-nm sample using 15-keV muons. Spectra are offset vertically for clarity. (b) Amplitude of the oscillating component of the wTF spectra extracted from least-squares refinements using the 70-nm sample with 3 different muon energies. The data shown were collected in a warming sequence. Error bars represent the propagated ESDs of the refined parameters determined by the fits. }
	
\end{figure}

\textbf{\textit{Confirming the lack of depth dependence of the magnetism.}}
Representative weak transverse field (wTF) \musr\ spectra for the 300-nm film with 15-keV muons are shown in Fig.~S\ref{fig:wtf}(a). The reduction of the oscillating amplitude as the temperature is lowered reflects the development of static magnetism in an increasingly large fraction of the sample volume probed by the muons. To determine whether there is any depth dependence of the magnetism, the 70-nm film was studied with muons of 4, 8, and 11~keV, corresponding to mean stopping depths of $20 \pm 20$~nm, $40 \pm 20$~nm, and $55 \pm 20$~nm. For the ZF measurements, there was no difference in the refined oscillation frequency measured with different muon energies outside of the statistical uncertainty of the fit parameters. Likewise, the wTF measurements revealed essentially identical temperature evolution of the asymmetry spectra. We display the wTF oscillation amplitudes as a function of temperature in Fig.~S\ref{fig:wtf}(b). The data collected from all muon energies overlap with each other very closely, indicating that there is little or no depth dependence of the magnetic phase transition in the film.

\begin{figure}
	
	\includegraphics[width=70mm]{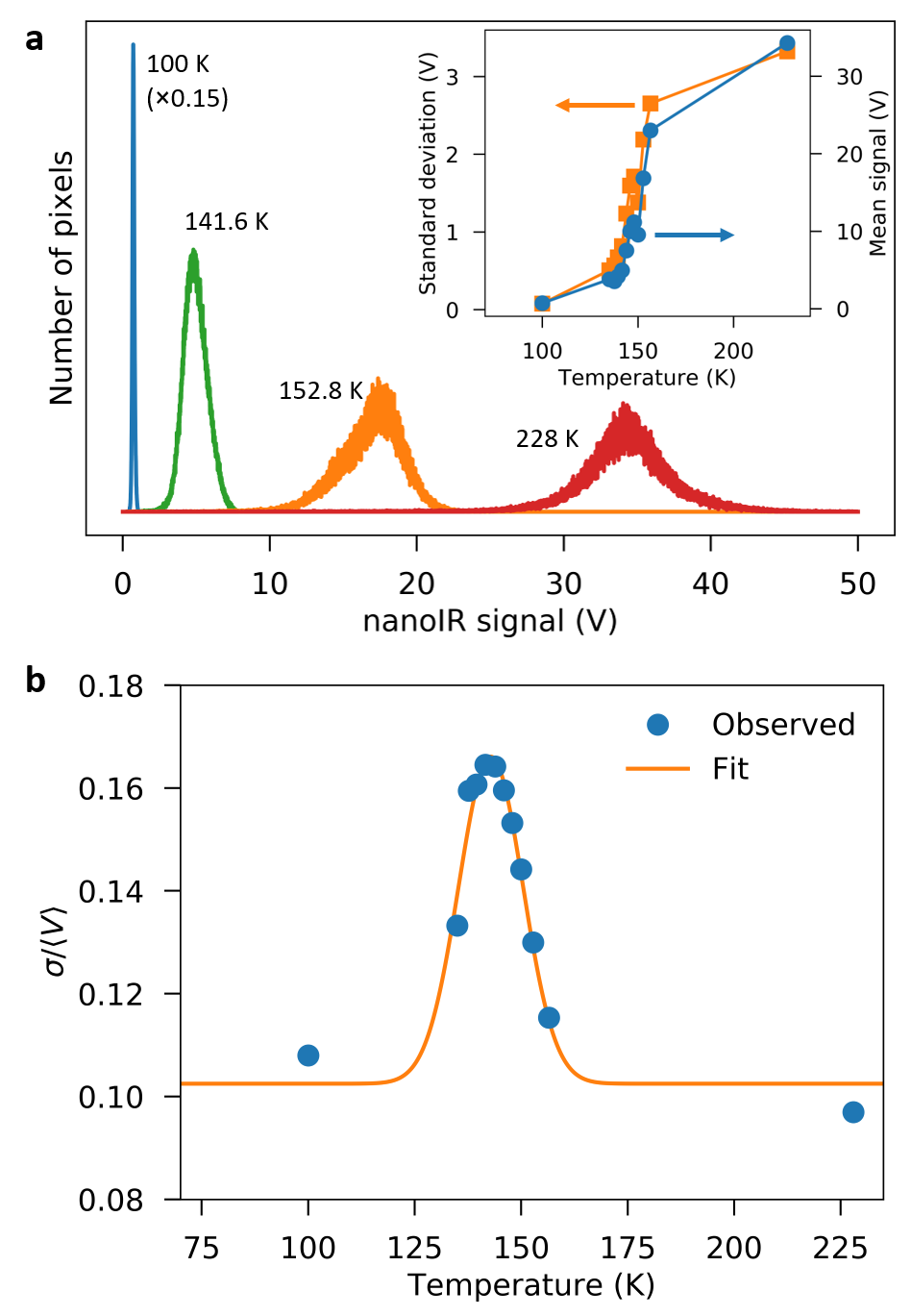}
	\caption{\label{fig:histograms} (Color online) (a) Histograms of the nanoIR signal in each pixel for images collected from the 70-nm \vo\ sample at four representative temperatures. Inset: Temperature dependence of the standard deviation (orange squares) and mean (blue circles) of the nanoIR histograms. (b) The relative width, $\sigma/\langle V \rangle$, of the nanoIR histograms for all measured temperatures (blue circles), along with a Gaussian fit (orange curve). The horizontal axis displays the nominal temperature recorded for each measurement, but the lack of cross-instrument thermometry calibration makes quantitative comparison unreliable.}
\end{figure}

\textbf{\textit{Nano-IR characterization.}}
To examine the MIT in detail, we performed nanoIR measurements of the 70-nm sample at 12 temperatures spanning the transition. Additional details can be found elsewhere~\cite{mcleo;np16}. The data were collected in a cooling sequence. To estimate the metallic and insulating phase fractions, we first generated histograms of the nanoIR voltage measured in each pixel for each image taken. The histograms displayed in Fig.~S\ref{fig:histograms}(a) were collected at four representative nominal temperatures: 228~K, well above the MIT; 152.8~K and 141.6~K, expected to be in the phase-separated transition region between metal and insulator; and 100~K, which is fully insulating.
The dominant trend in the histograms as the temperature is lowered is an abrupt reduction of the mean value and standard deviation of the nanoIR signal distribution. The inset in Fig.~S\ref{fig:histograms}(a) shows the standard deviation (left vertical axis) and mean (right vertical axis) of each histogram as a function of temperature, with a clear crossover region observed from about 140~K to 160~K. 

\begin{figure}
	
	\includegraphics[width=70mm]{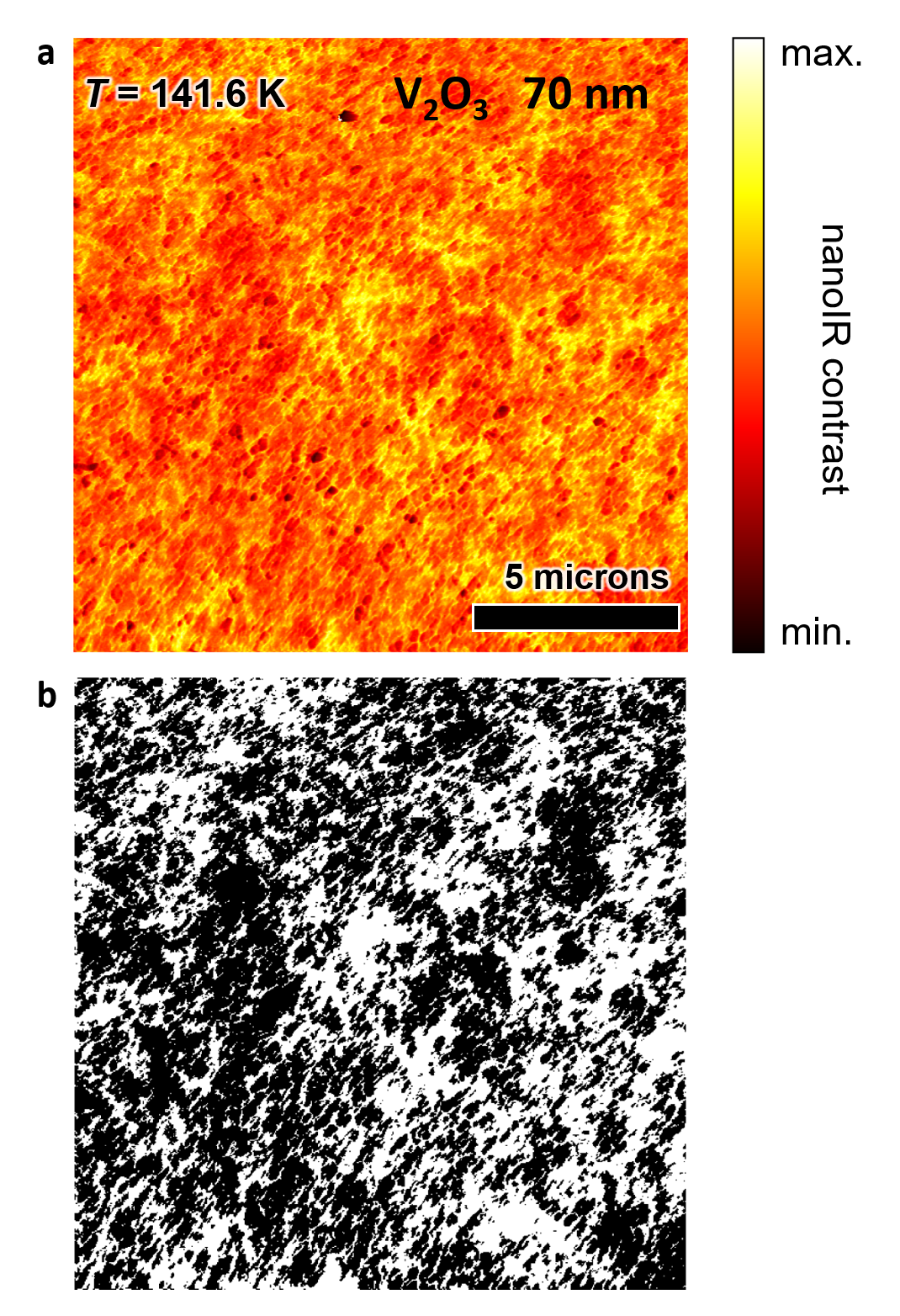}
	\caption{\label{fig:phasefrac} (a) Spatially resolved infrared spectroscopy measurements of the 70-nm sample of V2O3 at a nominal temperature of 141.6 K showing spatial phase separation between metallic (light-colored) and insulating (dark-colored) domains. (b) Same as (a), but binarized such that all pixels with a nanoIR voltage above a certain numerical threshold are white (representing metallic regions) and all pixels below the threshold are black (insulating regions).}
	
\end{figure}

The relative width of each histogram, which we define as $\sigma/\langle V \rangle$ with $\sigma$ being the standard deviation and $\langle V \rangle$ the mean value for a given nanoIR image, is expected to be largest (smallest) for the images with maximal (minimal) inhomogeneity. In other words, images collected at temperatures in the middle of the transition, where the sample is segregated into metallic and insulating regions, should display a greater relative width than those collected at temperatures well above or below the transition. This expectation is borne out by the data, as shown by the blue circles in Fig.~S\ref{fig:histograms}(b). We observe a clear peak centered around 143~K, which can be well fit by a function consisting of a Gaussian and a constant offset (orange curve in the figure). This systematic temperature evolution of the relative width reflects the progress of the electronic transition, allowing us to estimate the temperature-dependent insulating phase fraction $f_{\mathrm{ins}}(T)$ as 
\begin{align}\label{eq:insFrac}
	f_{\mathrm{ins}}(T) = \frac{\int_{T}^{\infty} A \exp{\left[-B(T'-T_0)^2\right]}\mathrm{d}T'}{\int_{-\infty}^{\infty} A \exp{\left[-B(T'-T_0)^2\right]}\mathrm{d}T'},
\end{align}
where $A \exp{\left[-B(T'-T_0)^2\right]}$ is the Guassian function determined from the fit to the data in Fig.~S\ref{fig:histograms}(b). To evaluate the reliability of this method, we can take the resulting insulating fraction for a given temperature (we will consider 141.6~K as an example), determine the numerical threshold separating insulating and metallic regions that yields the same insulating fraction, and generate a binarized version of the nanoIR image for comparison with the original image. This is done in Fig.~S\ref{fig:phasefrac}(b), where white (black) regions are greater (less) than the threshold, representing metallic (insulating) regions. This binarized image corresponds nicely to the full image shown in Fig.~S\ref{fig:phasefrac}(a). We note that this method of estimating the insulating phase fraction differs from that reported in Ref.~\onlinecite{mcleo;np16}. In that case, the distributions were clearly bimodal, allowing a two-component fit to the data to obtain the phase fractions directly. In this case, distinct peaks were not as easily resolved, so the two-component model failed to produce physically meaningful results.